\documentclass[12pt]{article}

\catcode`\@=11
\def\@normalsize{\@setsize\normalsize{15pt}\xiipt\@xiipt
\abovedisplayskip 14pt plus3pt minus3pt%
\belowdisplayskip \abovedisplayskip
\abovedisplayshortskip  \z@ plus3pt%
\belowdisplayshortskip  7pt plus3.5pt minus0pt}
\def\small{\@setsize\small{13.6pt}\xipt\@xipt
\abovedisplayskip 13pt plus3pt minus3pt%
\belowdisplayskip \abovedisplayskip
\abovedisplayshortskip  \z@ plus3pt%
\belowdisplayshortskip  7pt plus3.5pt minus0pt
\def\@listi{\parsep 4.5pt plus 2pt minus 1pt
            \itemsep \parsep
            \topsep 9pt plus 3pt minus 3pt}}
\def\underline#1{\relax\ifmmode\@@underline#1\else
        $\@@underline{\hbox{#1}}$\relax\fi}
\@twosidetrue \relax \catcode`@=12
\catcode`\@=11

\def\section{\@startsection{section}{1}{\z@}{3.5ex plus 1ex minus
   .2ex}{2.3ex plus .2ex}{\large\bf}}

\def\ps@headings{\def\@oddfoot{}\def\@evenfoot{}
\def\@oddhead{\hbox{}\hfill
        \makebox[.5\textwidth]{\raggedright\ignorespaces --\thepage{}--
        \hfill }}
\def\@evenhead{\@oddhead}
\def\subsectionmark##1{\markboth{##1}{}}
} \ps@headings \catcode`\@=12 \relax

\def\figcap{\section*{Figure Captions\markboth
        {FIGURECAPTIONS}{FIGURECAPTIONS}}\list
        {Fig. \arabic{enumi}:\hfill}{\settowidth\labelwidth{Fig. 999:}
        \leftmargin\labelwidth
        \advance\leftmargin\labelsep\usecounter{enumi}}}
 \relax
\def\tablecap{\section*{Table Captions\markboth
        {TABLECAPTIONS}{TABLECAPTIONS}}\list
        {Table \arabic{enumi}:\hfill}{\settowidth\labelwidth{Table 999:}
        \leftmargin\labelwidth
        \advance\leftmargin\labelsep\usecounter{enumi}}}
 \relax
\def\reflist{\section*{References\markboth
        {REFLIST}{REFLIST}}\list
        {[\arabic{enumi}]\hfill}{\settowidth\labelwidth{[999]}
        \leftmargin\labelwidth
        \advance\leftmargin\labelsep\usecounter{enumi}}}
 \relax
\catcode`\@=11
\def\marginnote#1{}
\newcount\hour
\newcount\minute
\newtoks\amorpm
\hour=\time\divide\hour by60 \minute=\time{\multiply\hour by60
\global\advance\minute by- \hour}
\edef\standardtime{{\ifnum\hour<12 \global\amorpm={am}%
    \else\global\amorpm={pm}\advance\hour by-12 \fi
    \ifnum\hour=0 \hour=12 \fi
    \number\hour:\ifnum\minute<100\fi\number\minute\the\amorpm}}
\edef\militarytime{\number\hour:\ifnum\minute<100\fi\number\minute}

\def\draftlabel#1{{\@bsphack\if@filesw {\let\thepage\relax
  \xdef\@gtempa{\write\@auxout{\string
    \newlabel{#1}{{\@currentlabel}{\thepage}}}}}\@gtempa
    \if@nobreak \ifvmode\nobreak\fi\fi\fi\@esphack}
     \gdef\@eqnlabel{#1}}
\def\@eqnlabel{}
\def\@vacuum{}
\def\draftmarginnote#1{\marginpar{\raggedright\scriptsize\tt#1}}
\def\draft{\oddsidemargin -.5truein
        \def\@oddfoot{\sl preliminary draft \hfil
        \rm\thepage\hfil\sl\today\quad\militarytime}
        \let\@evenfoot\@oddfoot \overfullrule 3pt
        \let\label=\draftlabel
        \let\marginnote=\draftmarginnote
\def\@eqnnum{(\theequation)\rlap{\kern\marginparsep\tt\@eqnlabel}%
\global\let\@eqnlabel\@vacuum}  }
\def\preprint{\twocolumn\sloppy\flushbottom\parindent 1em
        \leftmargini 2em\leftmarginv .5em\leftmarginvi .5em
        \oddsidemargin -.5in    \evensidemargin -.5in
        \columnsep 15mm \footheight 0pt
        \textwidth 250mmin      \topmargin  -.4in
        \headheight 12pt \topskip .4in
        \textheight 175mm
        \footskip 0pt
\def\@oddhead{\thepage\hfil\addtocounter{page}{1}\thepage}
        \let\@evenhead\@oddhead \def\@oddfoot{} \def\@evenfoot{}
}
\def\titlepage{\@restonecolfalse\if@twocolumn\@restonecoltrue\onecolumn
     \else \newpage \fi \thispagestyle{empty}\c@page\z@
        \def\thefootnote{\fnsymbol{footnote}} }
\def\endtitlepage{\if@restonecol\twocolumn \else  \fi
        \def\thefootnote{\arabic{footnote}}
        \setcounter{footnote}{0}}  
\catcode`@=12 \relax

\def\ps@headings{\def\@oddfoot{}\def\@evenfoot{}
\def\@oddhead{\hbox{}\hfill
        \makebox[.5\textwidth]{\raggedright\ignorespaces --\thepage{}--
        \hfill }}
\def\@evenhead{\@oddhead}
\def\subsectionmark##1{\markboth{##1}{}}
} \ps@headings \relax
\newcommand{\newc}{\newcommand}

\newc{\ra}{\rightarrow}
\newc{\lra}{\leftrightarrow}
\newcommand{\ba}{\begin{eqnarray}}
\newcommand{\ea}{\end{eqnarray}}

\def\eps{\epsilon}

\usepackage{epsfig}

\begin{document}
\def\firstpage#1#2#3#4#5#6{
\begin{titlepage}
\nopagebreak
\title{
\vfill {#3}}
\author{\large #4 \\[1.0cm] #5}
\maketitle \vskip -7mm \nopagebreak
\begin{abstract}
{\noindent #6}
\end{abstract}
\vfill
\begin{flushleft}
\rule{16.1cm}{0.2mm}\\[-3mm]
\end{flushleft}
\thispagestyle{empty}
\end{titlepage}}
\def\simlt{\stackrel{<}{{}_\sim}}
\def\simgt{\stackrel{>}{{}_\sim}}
\date{}
\firstpage{3118}{IC/95/34} {\large\bf
A D-brane inspired   Trinification model.}
 {G. K. Leontaris${}^{*}$ and J. Rizos}
{\normalsize\sl Theoretical Physics Division, Ioannina University,
GR-45110 Ioannina, Greece\\ [2.5mm]
 }
{We describe the basic features  of model building in the context
of intersecting D-branes. As an example, a D-brane inspired construction
with $U(3)_C\times U(3)_L\times U(3)_R$ gauge symmetry is proposed -which
is the analogue of the Trinification model- where the unification porperties
and some low energy implications on the fermion masses are analysed. }

\vfill {\it ${}^{*}$Talk presented at the ``Corfu Summer
Institute'', Corfu-Greece,  September 4-14, 2005}
\newpage
\section{Introduction}
Model building establishes the connection between the mathematical
formulation of a physics theory and the known (experimentally
discovered) world of elementary particles. In High Energy Physics
the ``known world'' is defined as the one which is described by
the Standard Model (SM). The SM spectrum, consists of three
flavors of LH lepton doublets and quarks, the RH electrons, up and
down quarks respectively plus gauge bosons and the Higgs field
(although not yet discovered).

During the last three decades we have learned a lot from the
attempts to extend the SM. As early as 1974, the observation that
gauge couplings converge at large scales, suggested unification of
the three forces at a scale $M_U\sim 10^{15}$ GeV. This
observation led to the incorporation of the SM into a gauge group
with higher symmetry i.e., $SU(5)$ etc (Grand Unification).  A
number of Grand Unified Theories (GUTs) (Pati-Salam model,  SO(10)
GUT etc )\footnote{For an early review see
\cite{Langacker:1980js}}
 predicted also the existence of the RH neutrino. Its existence
can lead to a light Majorana mass (through the see-saw mechanism)
which is now confirmed by the present experimental data. Next, the
incorporation of supersymmetry~\cite{Nilles:1983ge} into the game of
unification had a big success: This was the solution of the
hierarchy, which was a fatal problem in all non-supersymmetric
GUTs. However, the cost to pay was the doubling of the spectrum,
with the inclusion of superpartners and many arbitrary parameters.
For example, the number of arbitrary parameters one counts in the
Minimal Supersymmetric Standard Model (MSSM) are pretty much above
100. String theory model building went a bit further: For
instance, one could calculate from the first principles of the
theory the Yukawa couplings~\cite{Leontaris:1999ce}.
Non-renormalizable contributions of the form $\eps^n Q u^c h$,
$\eps=\frac{\langle\phi\rangle}{M_S}$, (and similar terms for the
other masses), where $\phi$ is a singlet and $M_S$ the string
scale, were also calculated to any order. This gave the
possibility to determine the structure of the fermion mass
matrices in terms of a few known expansion
parameters\cite{Ibanez:1994ig}.

This kind of mass textures gives rise to successful hierarchies of
the form $m_u:m_c:m_t$ $\sim \eps^6:\eps^2:1$ up to order one
coefficients (and similarly for the other fermions). A number of
other phenomenological problems however, were not resolved even in
the string context. For example, there was no systematic way of
eliminating baryon violating operators of any dimension, since,
even if they are absent at  the three level (due to the existence
of a possible symmetry), they generally appear in higher order
corrections. Another aesthetically and experimentally unpleasant
fact is that the string scale which, in all  models obtained from
the heterotic string theory, is of the order of the Planck
scale~\cite{Kaplunovsky:1987rp}. Yet, the non-discovery
of the superpartners at energies expected to be there, is also
another  headache.

It now appears  that many of the above unanswered questions and
puzzles might have a solution in models built in the context of
branes immersed in higher dimensions~\cite{Polchinski:1996na}.
Indeed, the models built in this context offer possibilities for
solving the above problems: a class of them  allows a low
unification scale of the order of a few TeV~\cite{Bachas:1998kr}, therefore
supersymmetry is not necessary since there is no hierarchy
problem; further, in certain cases, (type I string theory) the
presence of internal magnetic fields\cite{Bachas:1995ik} provides
a concrete realization of split supersymmetry~\cite{Antoniadis:2004dt},
therefore, intermediate or higher string scales are also possible
in this case\cite{split}. Further, the gauge group structure
obtained in these models contains $U(1)$ global symmetries, one of
them associated with the baryon number, so that baryon number
violation is prohibited to all orders in perturbation theory.

\section{Intersecting branes}

In the construction of D-brane models  the basic ingredient is the
brane stack, i.e. a certain number of parallel, almost coincident
D-branes.  A single D-brane carries a $U(1)$ gauge symmetry which
is the result of the reduction of the ten-dimensional Yang-Mills
theory. A stack of $N$ parallel branes gives rise to a $U(N)$
gauge theory where the gauge bosons correspond to open strings
having both their ends attached to some of the branes of the
various stacks. The compact space is taken to be  a
six-dimensional torus $T^6=T^2\times T^2\times T^2$, however, for
simplicity,  let us assume first the intersections on a single
$T^2$. $D$-branes in flat space lead to non-chiral matter.
Chirality arises when they are wrapped on a torus. In this case
chiral fermions sit in singular points in a transverse space while
the number of fermion
generations~\cite{Blumenhagen:2000wh,Aldazabal:2001cn}, and other
fermions in intersecting branes are related to the two distinct
numbers of wrappings of the branes around the two circles $R_1,
R_2$ of the torus~\footnote{This picture is the dual of D-branes
with magnetic flux~\cite{Bachas:1995ik}}.  Consider intersecting
D6-branes filling the four-dimensional space-time with $(n,m)$
wrapping numbers. (This means that we wrap $n$-times the brane
along the circle of $R_1$ radius and $m$-times along $R_2$.) For
two stacks $N_a, N_b$, we denote with $(n_a,m_a)$ and $(n_b,m_b)$
the wrappings respectively. The gauge group is $U(N_a)\times
U(N_b)$ while the fermions (which live in the intersections)
belong to the bi-fundamental $(N_a,\bar N_b)$, (or $(\bar N_a,
N_b)$). We may also obtain representations in $(N_a,N_{b^*})$ from strings
attached on the branes $a,b^*$, where $b^*$ is the mirror of the $b$-brane
under the $\Omega\,R$ operation, $\Omega$ beeing the whorld-sheet
parity operation and $R$ a geometrical action.
 The number of the intersections on the two-torus is given
by $I_{ab}=n_am_b-m_an_b$, for $(N,\bar N)$ and $I_{ab^*}=n_am_b+m_an_b$
 for $(N_a,N_{b^*})$. These also equal the number
of the chiral fermion representations obtained at the
intersections. Additional pairs can be constructed by the action
on the vector $\vec v=(n,m)$ with the $SL(2,Z)$. The latter
preserves the intersection numbers, therefore it preserves also
the number of chiral fermions. The $SL(2,Z)$ elements act:
\ba
\left(\begin{array}{c}
n'\\m'\\\end{array}\right)&=&\left(\begin{array}{cc}
a&b\\c&d\\\end{array}\right)\,\left(\begin{array}{c}n\\m
\\\end{array}\right)
\ea
(Since $ad-bc=1$, it follows $n_a'm_b'-m_a'n_b'=n_am_b-m_an_b$.)

The gauge couplings of the theory are given as follows: Let $g$ be
the metric on the torus
\ba
g&=&(2\pi)^2\left(\begin{array}{cc}
      R_1^2&R_1R_2\cos\theta\\
      R_1R_2\cos\theta&R_2^2\\
      \end{array}
\right) \label{2tmetric}
\ea
where $\theta$ is the angle of the two vectors defining the torus
lattice. The length of the $\vec v=(n,m)$ wrapping, is then
$\ell_{nm}=\sqrt{g_{ab}v^av^b}$, i.e.,
\ba
\ell_{mn}&=&2\pi\sqrt{n^2R_1^2+m^2R_2^2+2nmR_1R_2\cos\theta}
\label{lengthnm}
\ea
The gauge coupling $g_a$ of the $a^{th}$ group is  given by
\ba
\frac{4\pi^2}{g_a^2}&=&\frac{M_S}{\lambda_{II}}\ell_{m_an_a}
\ea
where $M_S$ is the string scale, $\lambda_{II}$ is the type-II string coupling and
$\ell_{m_an_a}$ is given by (\ref{lengthnm}).
Yukawa coefficients are also calculable in these constructions in
terms of geometric quantities (area) of the
torus\cite{Blumenhagen:2000wh,Aldazabal:2001cn}. For example,
the size of the Yukawa coupling $y_{ijk}$ for a square torus is
\ba
y_{ijk}&=& e^{-\frac{R_1R_2}{\alpha'}A_{ijk}}
\ea
where $A_{ijk}$ is the area of the world-sheet connecting three vertices,
scaled by the area of the torus.

We may generalize these results, considering compactifications on
a 6-torus factorized as $T^6=T^2\times T^2\times T^2$. For
example, if we denote $(n_a^i,m_a^i)$ the wrapping numbers of the
D6$_a$ brane around the $i^{th}$ torus, the number of
intersections is given by the product of the intersections in each
of them
\ba
I_{ab}&=&\prod_{i=1}^3(n_a^im_b^i-n_b^im_a^i), \;\; {\rm for} \;(N_a,\bar N_b)
\\
I_{ab^*}&=&\prod_{i=1}^3(n_a^im_b^i+n_b^im_a^i), \;\; {\rm for} \;(N_a, N_b^*)
\ea
while cancellation of the $U(N)$ anomalies requires that the spectrum
should satisfy $\sum_b\,I_{ab}N_b=0$. To satisfy this critirion, one usualy has
to add additional matter states. For example, the fulfilment of this requirement
in deriving the Standard Model in~\cite{Ibanez:2001nd} led to the introduction of
the right-handed neutrinos.
We should note that further consistency conditions, as the cancellation of
RR-tadpoles for theories~\cite{Blumenhagen:2000wh} with open string sectors should be
satisfied by $n_a^i,m_a^i$.\footnote{See for example the recent
review\cite{Uraga} and references therein.}

Due to the fact that D-brane constructions generate $U(N)$ symmetries, from
$U(N)_a\ra SU(N)_a\times U(1)_a$, we conclude that several $U(1)$ factors appear
in these models. For, example, the derivation of the SM may be obtained from
a set of $U(3), U(2)$ and several $U(1)$ brane stacks, leading to
a symmetry\cite{Antoniadis:2002en,Gioutsos:2005uw}
\ba
{U(3)}_C\times{U(2)}_L\times\prod_{i=1}^n{U(1)}^i=
{SU(3)}_C\times{SU(2)}_L\times U(1)_C\times U(1)_L\times\prod_{i=1}^n{U(1)}^i \label{ggg}
\ea
Fermion and Higgs representations are charged under $U(1)_{C,L,i}$.
Some of these $U(1)$ symmetries, play a particular role in the low energy effective theory.
 For example, $U(1)_C$ is related to
baryon number, since all quarks carry the same $U(1)_C$-charge. Thus, all global symmetries
of SM are gauge symmetries in the context of D-brane constructions. Further, the $U(1)$ factors
have mixed anomalies with the non-abelian groups $SU(N_a)$ given by $A_{ab}=(I_{ab}-I_{a^*b})N_a/2$.
It happens that one linear combination remains anomaly free and contributes to the hypercharge
generator which in general is a linear combination of the form
\ba
Q_Y&=&\sum_m^{n+2} c_m\,Q_m
\ea
where the $c_m$ are to be specified in terms of the hypercharge assignment of the
particle spectrum of a given D-brane construction.
 The remaining $U(1)$ combinations carry anomalies which are cancelled by
a generalized Green-Schwartz mechanism and the corresponding gauge bosons become massive.
These symmetries however, persist in the low-energy theory as global symmetries.

 \begin{figure}[h]
\centering
\includegraphics[scale=.4]{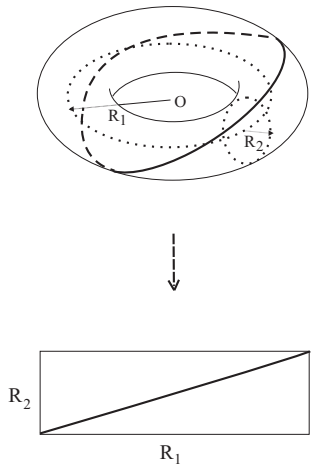}
\caption{Schematic representation of a $(1,1)$  D-brane wrapping on a $T^2$ torus.}
 \label{t2w}
\end{figure}

\section{A $U(3)^3$ brane inspired model}

We now  present  a specific non-supersymmetric example with gauge
symmetry $U(3)^3$ which can be considered as the analogue of the
``Trinification'' model proposed long time
ago~\cite{Glashow:1984gc,Rizov:1981dp}\footnote{For further
explorations, as well as supersymmetric and string versions of the
Trinification model see\cite{Babu:1986gi}-\cite{Choi:2003ag}} .
We will therefore describe here the steps one has to follow for a
viable D-brane construction. To generate this group one needs
three stacks of D-branes, each stack containing 3 parallel almost
coincident branes in order to form the $U(3)$ symmetry. We write
the complete gauge symmetry as
$$ U(3)_C\times U(3)_L\times U(3)_R,$$
so that the first $U(3)$ is related to  $SU(3)$ color, the second
involves the weak $SU(2)_L$ and the third is related to a possible
intermediate $SU(2)_R$ gauge group. Since $U(3)\ra SU(3)\times
U(1)$, we conclude that, in addition to the $SU(3)^3$ gauge group,
the D-brane construction  contains also three extra $U(1)$ abelian
symmetries. The $U(1)$ symmetry obtained from the color $U(3)_C\ra
SU(3)_C\times U(1)_C$ is related to the baryon
number~\cite{Antoniadis:2002en}. All baryons have the same charge
under $U(1)_C$ and consequently this $U(1)$ is identified with a
gauged baryon number symmetry. There are two more  abelian factors
from the chains $U(3)_L\ra SU(3)_L\times U(1)_L$, $U(3)_R\ra
SU(3)_R\times U(1)_R$ so the final symmetry can be written
\ba
SU(3)_c\times SU(3)_L\times SU(3)_R\times U(1)_C\times
U(1)_L\times U(1)_R\label{333111}
\ea
The abelian $U(1)_{C,L,R}$ factors have mixed anomalies with the
non-abelian $SU(3)^3$ gauge part with are determined by the
contributions of three fermion generations. There is an anomaly
free combination, namely~\cite{Leontaris:2005ax}
\ba
U(1)_{{\cal Z}'}&=& U(1)_C+U(1)_L+U(1)_R\label{aas}
\ea
which contributes to the hypercharge, while the two remaining combinations are
anomalous; these anomalies are cancelled by a generalized Green-Schwarz mechanism.
 \begin{figure}[h]
\centering
\includegraphics[scale=.9]{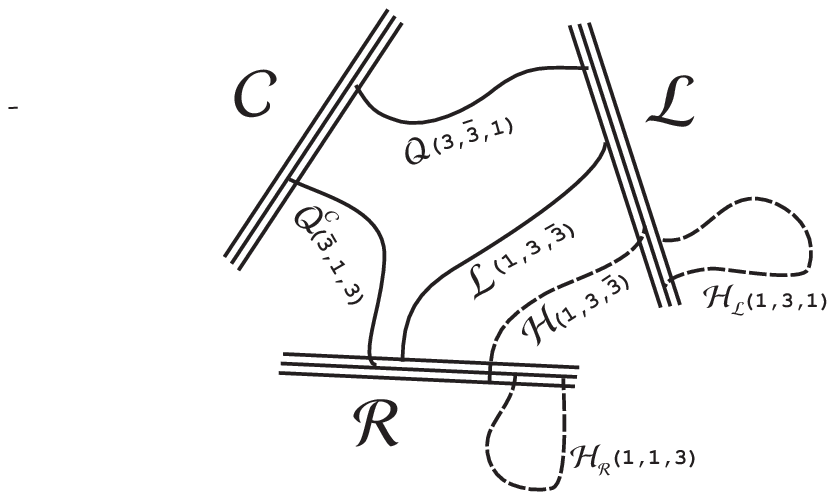}
\caption{Schematic representation of a $U(3)_C\times U(3)_L\times
U(3)_R$  D-brane configuration and the matter fields of the
model.}
 \label{u333}
\end{figure}
The possible representations which arise in this scenario should
accommodate the standard model particles and the necessary Higgs
fields to break the $U(3)^3$ symmetry down to SM. The spectrum of
a D-brane model involves two kinds of representations, those
obtained when the two string ends are attached to two different
branes and those whose both ends are on the same brane stack.  In
figure \ref{u333} we show the minimum number of irreps required to
accommodate the fermions and appropriate Higgs fields.  Under
(\ref{333111}) these states obtained from strings attached to two
different branes have the following quantum
numbers\footnote{ A schematic representation of the intersections of a $T^2$ torus which
 result to three fermion families  is shown in figure \ref{u333}, however,
in  a realistic scenario one should solve the complete system of equations
 for all states arising in this constructions on $T^2\times T^2\times T^2$.}
\ba
{\cal Q^{\hphantom{c}}}&=&(3,\bar 3,1)_{(+1,-1,\hphantom{+}0)}\label{QL}\\
{\cal Q}^c&=&(\bar 3,1,3)_{(-1,\hphantom{+}0,+1)}\label{QR}\\
{\cal L}^{\hphantom{c}}&=&(1,3,\bar
3)_{(\hphantom{+}0,+1,-1)}\label{Le}\\
{\cal H}^{\hphantom{c}}&=&(1,3,\bar
3)_{(\hphantom{+}0,+1,-1)}\label{Hi}
\ea
while the states arising from strings with both ends on the same
3-stack are
\ba
{\cal H}_{\cal L}&=&(1,3,1)_{(0,-2,0)}\label{HL}
\\
{\cal H}_{\cal R}&=&(1,1,3)_{(0,0,-2)}\label{HR}
\ea
Under  $ SU(3)_L\times SU(3)_R$ $\ra$
$\left[SU(2)_L\times
U(1)_{L'}\right]\,\times\,\left[U(1)_{R'}\times
U(1)_{\Omega}\right] $ and the $U(1)_{{\cal Z}'}$, we employ  the hypercharge embedding
\ba
Y=-\frac{1}{6} X_{L'}+\frac 13 X_{R'}+\frac 16{\cal Z}'\label{HC}
\ea
where $X_{L'}, X_{R'},{\cal Z}'$ represent the generators of the
corresponding $U(1)$ factors. Under the symmetry $SU(3)_C\times
SU(2)_L\times U(1)_Y\times U(1)_{\Omega}$ (\ref{QL}-\ref{HR})
decompose as follows
\ba
{\cal Q}^{\hphantom{c}}&=&
q\left(3,2;\frac{1}{6},0_{\Omega}\right)+g\left(3,1;-\frac{1}{3},0_{\Omega}\right)\label{QL1}\\
{\cal Q}^c&=&
d^c\left(\bar3,1;\frac{1}{3},1_{\Omega}\right)+u^c\left(\bar3,1;-\frac{2}{3},0_{\Omega}\right)
+g^c\left(\bar3,1;\frac{1}{3},-1_{\Omega}\right)\label{QR1}\\
{\cal L}^{\hphantom{c}} &=&
\ell^+\left(1,2;-\frac{1}{2},1_{\Omega}\right)+\ell^-\left(1,2;-\frac{1}{2},-1_{\Omega}\right)
+{\ell^c}\left(1,2;+\frac{1}{2},0_{\Omega}\right)\nonumber\\
&~&+\nu^{c+}(1,1;0,1_{\Omega})+\nu^{c-}(1,1;0,-1_{\Omega})+e^c(1,1;1,0_{\Omega})
\label{L1}
\\
{\cal H} &=&(1,3,\bar
3)=h^{d+}\left(1,2;-\frac{1}{2},1\right)+h^{d-}\left(1,2,-\frac{1}{2},-1\right)
+{h^u}\left(1,2;\frac{1}{2},0\right)\nonumber\\
&~&+{e_H^c}(1,1;1,0)+{\nu_{H}^{c+}}(1,1;0,1)+{\nu_{H}^{c-}}(1,1;0,-1),\;\;\;
\label{Hsm}
\\
{\cal H}_{\cal L} &=&(1,3,1)=\hat
h_L^{+}\left(1,2;-\frac{1}{2},0\right)+\hat\nu_{{\cal H}_{\cal
L}}\left(1,1;1,0\right)\label{hl}
\\
{\cal H}_{\cal R} &=&(1,1,3)= {\hat
e_H^c}(1,1;1,0)+{\hat\nu_{{\cal H}_{\cal
R}}^{c+}}(1,1;0,1)+\hat\nu_{{\cal H}_{\cal
R}}^{c-}(1,1;0,-1)\label{hr}
\ea
Representation (\ref{QL1}) includes the  left handed quark
doublets and an additional colored triplet with quantum numbers as
those of the down quark, while representation (\ref{QR1}) contains
the right-handed partners of (\ref{QL1}). Further, (\ref{L1})
 involves the lepton doublet, the right-handed electron and
its corresponding neutrino, two additional $SU(2)_L$ doublets and
another neutral state, called neutreto\cite{Glashow:1984gc}. The
Higgs sector consists of (\ref{Hsm}) which is the same
representation as that of the lepton fields, and the left and
right triplets (\ref{hl}) and ({\ref{hr}) respectively.
 \begin{figure}[h]
\centering
\includegraphics[scale=.6]{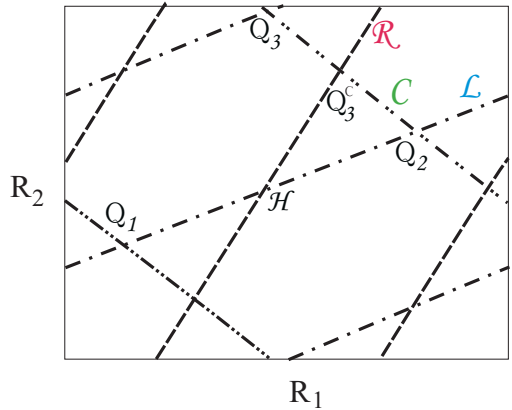}
\caption{Schematic representation of the intersections on a $T^2$
for a three-generation  $U(3)_C\times U(3)_L\times
U(3)_R$  D-brane
model.}
 \label{u333}
\end{figure}

\subsection{ Mass scales, Symmetry breaking and  Yukawa couplings}

\subsubsection{Mass scales} The reduction of the
${SU(3)}^3\times{U(1)}^3$ to the SM is in general associated with
three different scales corresponding to the $SU(3)_R$, $SU(3)_L$
and $U(1)_{{\cal Z}'}$ symmetry breaking. We will assume here for
simplicity that the $SU(3)_{L,R}$ and $U(1)_{{\cal Z}'}$
symmetries break simultaneously at a common scale $M_R$, hence the
model is characterized only by two large scales, the String/brane
scale $M_S$, and the  scale $M_R$. Clearly, the $M_R$ scale cannot
be higher than $M_S$, i.e., $M_R\le M_S$, and the equality holds
if the $SU(3)_R\times SU(3)_L$ symmetry breaks directly at $M_S$.
In a D-brane realization of the proposed model, since the three
$U(3)$ gauge factors  originate from 3-brane stacks that span
different directions of the higher dimensional space, the
corresponding  gauge couplings $\alpha_{C,L,R}$ are not
necessarily equal at the string scale $M_S$.  However, in certain
constructions, at least two D-brane stacks can be superposed and
the associated couplings are equal\cite{Antoniadis:2002en}. In our
bottom up approach, a crucial role in the determination of the
scales $M_{R,S}$ is played by the neutrino physics. More
precisely, in order to obtain the correct scale for the light
neutrino masses, which are obtained through a see-saw mechanism
and are found to be of the order $m_{\nu}\sim m_W^2/M_S$, the
string scale $M_S$ should be in the range $M_S\sim
10^{13}-10^{15}$ GeV.  In order to determine the range of
$M_S,M_R$, we use as inputs the low energy data for
$\alpha_3,\alpha_{em}$ and $\sin^2\theta_W$ and perform a one-loop
renormalization group analysis. The cases $\alpha_L=\alpha_R$ and
$\alpha_R=\alpha_C$ presented in Table \ref{ytab3a} are found to
be consistent with the neutrino data.
\begin{table}[!h]
\centering
\begin{tabular}{|c|c|l|}
\hline model&$M_R/GeV$ &${M_S}/{GeV}$\\
\hline
 $a_L=a_R$&$
 1.7\times 10^{9}$&$
 >1.7\times 10^{9}$\\
\hline $a_L=a_C$&$
<2.3\times 10^{16}$&$
>2.3\times 10^{16}$\\
 \hline
$a_C=a_R$&$
<2.3\times 10^{11}$&$
>2.3\times 10^{11}$\\
 \hline
\end{tabular}
\caption{\label{ytab3a} Upper and lower bounds for $SU(3)_R$
breaking scale ($M_R$)  and the corresponding String scale ($M_S$)
for the three cases $a_L=a_C$, $a_R=a_C$ and $a_L=a_R$. }
\end{table}
In particular, we find that the case $\alpha_L=\alpha_R$
predicts $M_R$  constant, i.e., independent of the common gauge
coupling $a\equiv\alpha_L=\alpha_R$ and $M_S$ also in the required
region. For $\alpha_R=\alpha_C$, we also obtain  $M_S\ge
2.3\times10^{11}$GeV.\footnote{The case $\alpha_L=\alpha_C$ is
ruled out by neutrino data, since it predicts $M_S> 10^{16}$GeV.}

\subsubsection{The Symmetry breaking}
 The Higgs states (\ref{Hsm}-{\ref{hr}) are sufficient to break the
original gauge symmetry $U(3)^3$ down to the Standard
Model\cite{Leontaris:2005ax}, however, according to
ref\cite{Glashow:1984gc}, a non-trivial KM mixing and quark mass
relations would require at least two Higgs fields in
$(1,3,\bar3)$.  We should mention however, that in string or
intersecting brane models, Yukawas are calculable in terms of
geometric quantities -such as torus area-  thus, from this point
of view  a second Higgs is not necessary. To break the symmetry
and provide with masses the various matter multiplets we  assume
two Higgs in $(1,3,\bar 3)$ and a pair ${\cal H}_{\cal
L}=(1,3,1)$, ${\cal H}_{\cal R}=(1,1,3)$ with the following vevs:
\ba
{\cal H}_1&\ra&\langle h^u_1  \rangle =u_1,\; \langle h^{d-}_{1}
\rangle =u_2,\;
                   \langle {{\nu^{c+}_{H,}}_1} \rangle =U,\nonumber\\
{\cal H}_2&\ra&\langle h^u_2 \rangle =v_1,\; \langle {h^{d-}}_2
\rangle =v_2,\; \langle {h^{d+}}_2 \rangle =v_3, \langle
{\nu^{c-}_{H}}_2 \rangle =V_1,\; \langle {\nu^{c+}_{H}}_2 \rangle
=V_2\nonumber\\
{\cal H}_{\cal L}&\ra&\langle\hat \nu_{H_L}^{}\rangle=A_L\nonumber\\
{\cal H}_{\cal R}&\ra&\langle\hat\nu_{H_R}^{}\rangle=A_R\nonumber
\ea
The vevs $U,V_{1,2}$ and $A_{L,R}$ are taken of the order  $M_R$,
while $u_{1,2}$ and $v_{1,2}$ are of the order of the electroweak
scale.

\subsubsection{Fermion masses}
 In the
present $U(3)^3$ construction, due to the existence of the
additional $U(1)_{C,L,R}$ symmetries,  the following Yukawa
coupling is present  at the tree-level Yukawa potential
\ba
\lambda_{Q,1}^{ij}{\cal Q}_i\,{\cal Q}_j^c\,{\cal H}_a,
\;\;{a=1,2}
\ea
 It can provide quark masses as well
masses for the extra triplets. For the up quarks
\ba
m_{uu^c}^{ij}&=&\lambda_{Q,1}^{ij}u_1+\lambda_{Q,2}^{ij}v_1\label{uqm}
\ea
For the down-type quarks $d_i,d_j^c, g_i, g_j^c$, we obtain a
$6\times 6$ down type quark matrix in flavour space, of the form
\ba
m_{d}&=&\left(\begin{array}{cc}
m_{dd^c}&M_{gd^c}\\
m_{dg^c}&M_{gg^c}\label{mdg}
\end{array}
\right)
\ea
where $m_{dd^c}=\lambda_{Q,1}^{ij}\,u_2+\lambda_{Q,2}^{ij}\,v_2$
and $m_{dg^c}=\lambda_{Q,2}^{ij}\,v_3$ are $3\times 3$ matrices with
entries of the electroweak scale, while $M_{gd^c}=\lambda_{Q,2}^{ij}\,V_1$,
$M_{gg^c}=\lambda_{Q,1}^{ij}\,U+\lambda_{Q,2}^{ij}\,V_2$ are of the order
$M_R$. The diagonalization of the non-symmetric mass matrix
(\ref{mdg}) will lead to a light $3\times 3$ mass matrix for the
down quarks and a heavy analogue of the order of the $SU(3)_R$
breaking scale.

The extra $U(1)_{C,L,R}$ factors do not allow for a tree-level
coupling for the lepton fields. The lowest order allowed leptonic
Yukawa terms arise at fourth order. These are
\ba
\frac{f_{ij}^{ab}}{M_S}\, {\cal H}_a^\dagger \,{\cal
H}_b^\dagger\,{\cal L}_i\,{\cal L}_j +
\frac{\zeta_{ij}}{M_S}\,{\cal H}_{\cal L} \,{\cal H}_{\cal
R}^\dagger\,{\cal L}_i\,{\cal L}_j
\ea
where $f_{ij}^{ab},\zeta_{ij}$ are order one Yukawa couplings, and
$a,b=1,2$. These terms  provide with  masses the charged leptons
suppressed by a factor $M_R/M_S$ compared to quark masses. Thus, a
natural quark-lepton hierarchy arises in this model. They further
imply light Majorana masses for the three neutrino species through
a see saw mechanism. All the remaining states (lepton like
doublets and neutral singlets) obtain masses of the order
$M_R^2/M_S$\cite{Leontaris:2005ax}.

\section{Conclusions}

In this talk, we have described the basic features  of model
building in the context of intersecting D-branes. As an example,
 we have analysed a D-brane analogue of the trinification model
 which can be generated by three separate
stacks of D-branes. Each of the three stacks is formed by three
identical branes, resulting to an $U(3)_C\times U(3)_L\times
U(3)_R$ gauge symmetry for the model.  Since $U(3)\ra SU(3)\times
U(1)$, this symmetry is equivalent to the standard $SU(3)^3$
trinification gauge group supplemented by three abelian factors
$U(1)_{C,L,R}$. The main characteristics of the model are:

 $\bullet$ The three
$U(1)$ factors define an unique anomaly-free combination
$U(1)_{{\cal Z}'}=U(1)_{C}+U(1)_L+U(1)_R$ as well as two other
anomalous combinations whose anomalies can be cancelled by a
generalized Green-Schwarz mechanism.

 $\bullet$
The Standard Model fermions are represented by strings attached to
two different brane-stacks and belong to $(3,\bar 3,1)+(\bar
3,1,3)+(1,3,\bar 3)$ representations as is the case of the
trinification model.

 $\bullet$
The scalar sector contains  Higgs fields in $(1,3,\bar 3)$ (which
is the same representation which accommodates the lepton fields),
as well as Higgs in (1,3,1) and (1,1,3) representations which can
arise from strings whose both ends are attached on the same brane
stack. The Higgs fields break the $SU(3)_L\times SU(3)_R$ part of
the gauge symmetry down to $U(1)_{em}$; they further provide a
natural quark-lepton hierarchy since quark masses are obtained
from tree-level couplings, while, due to the extra $U(1)$
symmetries, charged leptons are allowed to receive masses from
fourth order Yukawa terms.

 $\bullet$ The $SU(3)_R$ breaking scale is found to be $M_R>
 10^9$ GeV, while a string scale $M_S\sim 10^{13-15}$GeV is predicted
 which suppresses the light Majorana masses through a see-saw
 mechanism down to sub-eV range as required by neutrino
 physics.

\vfill

 {\bf Ackmowledgements}. {\it This research was co-funded by the European
 Union in the framework of the program "Pythagoras I" (no.~1705 project 23)
  of the "Operational Program for Education and Initial  Vocational Training"
  of the 3rd Community Support Framework of the Hellenic Ministry of Education,
 funded by 25\% from national sources and by 75\% from the European Social Fund (ESF).}

\end{document}